\documentclass[aps,pra,floatfix,showpacs,preprint]{revtex4-1}

\usepackage{graphicx}
\usepackage{subfigure}
\usepackage{amsmath, amsfonts, amssymb, bm}
\usepackage{nicefrac}

\usepackage{amstext}
\usepackage{mathrsfs}
\newcommand{\Eqref}[1]{Eq.~\eqref{#1}}
\newcommand{\Figref}[1]{Fig.~\ref{#1}}

\begin{document}
\title{Electron-beam dynamics in a strong laser field including quantum radiation reaction}
\author{N.~Neitz}
\author{A.~Di Piazza}
\affiliation{Max-Planck-Institut f\"ur Kernphysik, Saupfercheckweg 1, D-69117 Heidelberg, Germany}

\begin{abstract}
The evolution of an electron beam colliding head-on with a strong plane-wave field is investigated in the framework of strong-field QED including radiation-reaction effects due to photon emission. Employing a kinetic approach to describe the electron and the photon distribution it is shown that at a given total laser fluence the final electron distribution depends on the shape of the laser envelope and on the pulse duration, in contrast to the classical predictions of radiation reaction based on the Landau-Lifshitz equation. Finally, it is investigated how the pair-creation process leads to a nonlinear coupled evolution of the electrons in the beam, of the produced charged particles, and of the emitted photons. 
\end{abstract}

\maketitle

\section{Introduction}

The vast experimental progress in the generation of intense laser pulses and ultrarelativistic particle beams makes it essential to attain a profound understanding of the dynamics of charged particles in the presence of electromagnetic background fields. The investigation of the latter is not only interesting from a theoretical point of view but also crucial for experimental applications, e.g., in accelerator and plasma physics. Already in the realm of classical electrodynamics the problem occurs on how to calculate the trajectory of a charged particle in an external field, including the continuous loss of energy and momentum due to the emission of electromagnetic radiation \cite{Landau:1975, Spohn:2004,Rohrlich_b_2007}. This so-called radiation-reaction (RR) problem consists in determining an equation of motion of the charged particle, which self-consistently incorporates energy-momentum loss. Historically, the Lorentz-Abraham-Dirac (LAD) equation has been suggested and it can be obtained by eliminating the 
degrees of freedom of the electromagnetic field in the coupled system of Maxwell's and single-particle Lorentz equations \cite{Rohrlich_b_2007}. The LAD equation contains additional terms apart from the Lorentz-force, which are responsible for RR effects. One of these terms, however, depends on the time-derivative of the charge acceleration, which in turn leads to the existence of so-called runaway solutions, with the particle acceleration exponentially increasing with time even in absence of any driving force \cite{Landau:1975, Spohn:2004,Rohrlich_b_2007}. On the other hand, it has been first noticed in \cite{Landau:1975}, that in the realm of classical electrodynamics, i.e., if quantum effects, like the recoil in photon emission, are negligible, the LAD equation can be consistently approximated via the so-called Landau-Lifshitz (LL) equation, that avoids the above-mentioned inconsistencies (see also \cite{Spohn:2000,Rohrlich:2002,DiPiazza:2012rev}). Even 
though the LL equation provides a consistent description of RR in classical electrodynamics, the experimental verification of this equation is still missing. In the so-called classical radiation dominated regime the electron emits an average energy in a single laser cycle, which is comparable to its initial energy and RR effects dominate the electron dynamics \cite{Koga_2005,DiPiazza:2008}. However, the classical radiation dominated 
regime is rather hard to be entered with present technology. In \cite{Di_Piazza_2009} strong signatures of RR have been predicted to occur also below the classical radiation dominated regime, based on the analytical solution of the LL equation in an arbitrary plane-wave field found in \cite{DiPiazza:2008}. Alternative proposals to measure RR effects in the classical domain have been suggested in \cite{Harvey_2011,Thomas_2012,Heinzl_2013}.

Since classical electrodynamics is contained in the underlying theory of quantum electrodynamics, the understanding of the quantum origin of RR is of fundamental importance. As we have mentioned, RR in classical electrodynamics include all effects, which go beyond the Lorentz dynamics and which stem from the action of the electromagnetic field generated by the charged particle, an electron for definiteness, on the electron itself, when it is driven by a background electromagnetic field. Analogously, RR in QED includes all possible processes that may arise starting from a single, external-field-driven electron \cite{DiPiazza:2010mv,DiPiazza:2012rev,Ilderton_2013PLB,Ilderton_2013PRD}. Thus, the complete description of RR in the quantum regime can be achieved by the calculation of the full $S$-matrix, including effects like radiative corrections, multiple photon emission and pair creation following photon emission. This implies that in the full quantum regime, the problem of RR is intrinsically multiparticle. 
However, in the moderate quantum 
regime, in which quantum effects are not too large and electron-positron pair production is negligible (see Sec. \ref{KinApp} for further details), the process that mainly gives rise to RR effects is the incoherent emission of multiple photons. (Note, for example, that radiative corrections $\delta m^2$ to the square $m^2$ of the electron mass are negligible as they are of the
order $1\%$ of $m^2$. Indeed, such radiative corrections roughly scale as $\alpha m^2$ \cite{Ritus:1985}, where $\alpha=e^2/\hbar c\approx 1/137$ is the fine-structure constant, with $e<0$ being the electron charge.) This regime is then single-particle as in classical electrodynamics and it has been investigated in detail in \cite{DiPiazza:2010mv} in the case of a background plane wave. In \cite{DiPiazza:2010mv} the so-called quantum radiation dominated regime has also been introduced, where the electron emits on average more than one photon with substantial recoil already in one laser period. In the quantum radiation dominated regime the electron dynamics is dominated by both quantum and RR effects. However, a sufficient increase of the intensity of the background electromagnetic field and/or of the initial electron energy will let enter a new regime, where neglecting pair creation is no longer permissible. In the collision of a laser field and electrons, pair production may 
occur due to two different processes \cite{DiPiazza:2012rev} (see also for a recent review on pair-production processes \cite{Ruffini}): (i) the photons emitted by the incoming electrons reach sufficiently high energies allowing for laser-assisted electron-positron production and (ii) the direct pair production by electrons via the emission of virtual photons. These two channels are treated in a unified way in strong-field QED \cite{Hu_2010,Ilderton_2011} (see also \cite{King_2013}). In the presence of an electromagnetic wave of sufficient high intensity, the emission of the photon and the transformation of the photon into an electron-positron pair do not occur in the same formation region \cite{Baier:1998vh,King_2013}. Since the created charged particles are thereupon accelerated by the laser field, the produced pairs will emit further photons and, under certain circumstances, prime the formation of the so-called QED cascades \cite{Bell_2008,Fedotov_2010,Elkina:2010up,Sokolov:2010am,Nerush:2010fe}.

In the present paper we first study the interaction of an electron beam colliding head-on with a strong plane wave in the nonlinear moderate quantum regime and follow the macroscopic kinetic approach \cite{Neitz2013} (see Sec. \ref{KinApp}), which allows to take into account the multiple incoherent emission of high-energy photons by a distribution of electrons in the quantum regime. Alternative approaches are the microscopic approach employed in \cite{DiPiazza:2010mv} or the stochastic model of RR investigated in \cite{Blackburn:2014}. 
Since RR is, generally speaking, a dissipative effect (see also Ref. \cite{Tamburini_2014} for a setup, which allows to use RR effects to control the electron dynamics), we focus our attention in Sec. \ref{Shape} on the effects stemming from the pulse shape and duration of the laser field. As we will see, such effects can be exploited as an observable for testing the predictions of QED in the quantum radiation dominated regime. We will see that already at available laser intensities of the order of $10^{22}\;\text{W/$\text{cm}^2$}$, the final electron distribution strongly depends on 
the shape and on the duration of the pulse 
also at a given pulse fluence, whereas the classical dynamics based upon the LL equation would predict no dependence in this case. Further, the influence of the form and of the duration of the pulse shape at a given laser fluence on the photon spectrum emitted by the electron distribution is also investigated. In Sec. \ref{PairProSec} we extend the approach in \cite{Neitz2013} and we will also take into account the possibility that emitted photons then transform into electron-positron pairs. Therefore, we include the corresponding probabilities in our kinetic approach and study how the inclusion of 
the process of pair creation influences the evolution of the charged particle and of the photon distribution functions. Moreover, numerical simulations for different pulse durations at a given laser fluence will be presented for two different initial electron energy distributions, allowing us to identify two distinguishable scenarios, where the laser pulse duration has diverse effects on the number of produced pairs. In the appendix A we amend a step in the derivation of the kinetic equation given in the Supplemental Material of \cite{Neitz2013} (Eq. (\ref{Kinetic1}) here). Finally, appendix B contains an example showing numerically the equivalence of the kinetic approach proposed here and the previous microscopic approach developed in \cite{DiPiazza:2010mv}.

\section{Kinetic Approach}
\label{KinApp}
We consider the collision of an electron beam with a plane wave characterized by the electric field $\bm{E}(\varphi)=E_0f(\varphi)\hat{\bm{z}}$. Here, $E_0$ is the laser electric field amplitude and $f(\varphi)$ is an arbitrary function of the laser phase $\varphi=\omega_0(t-y)$ such that $|f(\varphi)|_{\text{max}}\leq1$, with $\omega_0$ being the laser central angular frequency (units with $\hbar=c=1$ are used throughout). In the case of a single electron with the initial four-momentum $p_0^\mu=(\varepsilon_0,\boldsymbol{p}_0)$, the quantity $p_{0,-}=\varepsilon_0-p_{0,y}$ plays an important role in the electron dynamics, due to the special dependence of the plane wave on the space-time coordinates. In fact, without taking into account radiation reaction, the light-cone components $p_-(\varphi)=\varepsilon(\varphi)-p_y(\varphi)$, $\boldsymbol{p}_\perp(\varphi)$, and $p_+(\varphi)=(\varepsilon(\varphi)+p_y(\varphi))/2$ of the four-momentum $p^{\mu}(\varphi)=(\varepsilon(\varphi),\bm{p}(\varphi))$ of an 
electron (mass $m$) in the presence of the mentioned plane wave are given by \cite{Landau:1975}
\begin{align}
\label{pMinusConst}
p_-(\varphi)&\equiv p_{0,-},\\
\label{pPerpEvolv}
\boldsymbol{p}_\perp(\varphi)&=\bm{p}_{0,\perp}-e\bm{A}(\varphi),\\
\label{pPlusEvolv}
p_+(\varphi)&=\frac{m^2+\bm{p}_\perp^2(\varphi)}{2p_{0,-}},
\end{align}
where we have chosen the initial phase $\varphi_i=0$. Further, we have introduced the four-vector potential in Lorentz gauge $A^\mu(\varphi)=(0,\boldsymbol{A}(\varphi))=(0,-E_0F(\varphi)\hat{\bm{z}}/\omega_0)$, with $F(\varphi)=\int_0^\varphi d \varphi'\,f(\varphi')$. In order to describe the radiation of an electron in a plane-wave field, we use the well-known single photon emission probability per unit of the laser phase $\varphi$ and per unit $u=k_-/(p_--k_-)$ \cite{Ritus:1985}
\begin{equation}
\label{ToniProb}
 \frac{dP_{p_-}}{d\varphi du}=\frac{\alpha}{\sqrt{3}\pi}\frac{m^2}{\omega_0p_-}\frac{1}{(1+u)^2}\left[\left(1+u+\frac{1}{1+u}  \right)
\text{K}_{\frac{2}{3}}\left( \frac{2u}{3\chi(\varphi,p_-)}\right)-\int_{2u/[3\chi(\varphi,p_-)]}^\infty dx\, \text{K}_{\frac{1}{3}}(x) \right],
\end{equation}
where we have introduced the variable $k_-=\omega-k_y$ for the emitted photon with four-momentum $k^\mu=(\omega,\boldsymbol{k})$, where $\text{K}_\nu(x)$ is the modified Bessel function of $\nu$th order. The symbol $\chi(\varphi,p_-)$ in Eq. (\ref{ToniProb}) indicates the phase-dependent quantum nonlinearity parameter, which measures the importance of quantum effects like photon recoil \cite{DiPiazza:2012rev}. In our case, this is given by $\chi(\varphi,p_-)=p_-|E(\varphi)|/mE_{\text{cr}}$, with $E(\varphi)=E_0f(\varphi)$ and with $E_\text{cr}=m^2/|e|=1.3\times 10^{16}\;\text{V/cm}$ being the critical field of QED \cite{DiPiazza:2012rev}. Here, the probability in Eq. (\ref{ToniProb}) is averaged over the initial electron spin and summed over the final electron spin and photon polarization.  Furthermore, \Eqref{ToniProb} is the 
photon emission probability in the case of a constant crossed field of amplitude $E$ with the substitution $E\to |E(\varphi)|$ and it is 
only valid in the quasi static approximation \cite{Ritus:1985}. In turn, this approximation is valid if the relativistic parameter $\xi=|e| E_0/m\omega_0$ of the plane wave is much larger than unity. In fact, if $\lambda_0=2\pi/\omega_0$ is the central wavelength of the plane wave and thus the typical distance over which the plane wave varies, the radiation formation length $l_0$ of the photon production process at $\xi\gg 1$ is $l_0=\lambda_0/\xi$ and it is much smaller than $\lambda_0$ \cite{Ritus:1985}.

In our approach, we describe the electron beam via an electron distribution $n_{e^-}(\varphi,p_-)$. It is always assumed that the electron distribution depends only on $\varphi$ and on $p_-$, i.e., the motion along the perpendicular directions $x$ and $z$ is neglected (see also Ref. \cite{Green_2013} for a recent investigation of the influence of RR effects on the electron transverse momentum). By employing Eq. (\ref{pPerpEvolv}) with the initial condition $\bm{p}_{0,\perp}=\bm{0}$, it can be seen that the modulus of the transverse momentum $|\bm{p}_{\perp}|$ has the upper limit $\sim m\xi$ (note that for a laser field $|F(\varphi)|\sim 1$) and, in turn, that $p_+$ does not exceed in order of magnitude the quantity $m^2\xi^2/2p_-$. Thus, we can safely neglect the momentum components $\bm{p}_{\perp}$ and $p_+$ in our considerations in the case of ultra-relativistic electron bunches colliding head-on with a laser beam in which $p_-^\ast\gg m\xi$ , where $p_-^{\ast}$ indicates the typical value of the quantity $p_-$ of the electron distribution. In the case, for example, of a typical energy $\varepsilon^\ast=1\,\text{GeV}$ ($p_-^{\ast}\approx2\,\text{GeV}$) and of an optical 
laser field of 
intensity $10^{23}\,\text{W/$\text{cm}^2$}$, with $\omega_0=1.55\,\text{eV}$, we obtain $m\xi\lesssim 78\,\text{MeV}$ and $p_+\lesssim 1.5\,\text{MeV}$, which well fulfill the conditions $|\bm{p}_{\perp}|\ll p_-^{\ast}$ and $p_+\ll p_-^{\ast}$ (we ensured that these conditions are fulfilled during the whole interaction of the electron beam with the plane wave). In general, electron-positron pairs are also produced in the collision of the electron beam and the laser field, because the electrons emit photons, which in turn can still interact with the laser field \cite{Hu_2010,Ilderton_2011}. The possibility of electron-positron pair creation will be included in the kinetic approach and its effects will be discussed in Sec. \ref{PairProSec}. However, if we assume that the 
typical value $\chi^\ast=p_-^\ast E_0/mE_\text{cr}$ of the quantum nonlinearity parameter does not largely exceed 
unity, we 
are allowed to neglect pair production for the moment. In fact, we recall that the probability of pair production contains an exponential damping factor $\exp(-8/3\varkappa^{\ast})
$ \cite{Ritus:1985}, where $\varkappa^\ast=k_-^\ast\chi^\ast/p_-^\ast$, with $k_-^\ast<p_-^\ast$ being the typical value of the quantity $k_-$ of the emitted photons (see \cite{Ritus:1985}). In this framework and by also neglecting radiative corrections, which are high-order in $\alpha$, the kinetic equations [see Eq. (20.1) in Ref. \cite{Baier:1998vh}, \cite{Sokolov:2010am} and Appendix A]
\begin{align}
\label{Kinetic1}
\frac{\partial n_{e^-}(\varphi,p_-)}{\partial\varphi}=&\int_{p_-}^\infty dp_{i,-}\,n_{e^-}(\varphi,p_{i,-})\frac{dP_{p_{i,-}}}{d\varphi dp_-}-n_{e^-}(\varphi,p_{-})\int_0^{p_-}dk_-\frac{dP_{p_{-}}}{d\varphi dk_-}\\
\label{Kinetic2}
\frac{\partial n_\gamma(\varphi,k_-)}{\partial\varphi}=&\int_{k_-}^\infty dp_{i,-}\, n_{e^-}(\varphi,p_{i,-})\frac{dP_{p_{i,-}}}{d\varphi dk_-},
\end{align}
with 
\begin{align}
\label{PTrans}
\frac{dP_{p_{i,-}}}{d\varphi dp_-}&=\left\vert\frac{du}{dp_-}\right\vert\left.\frac{dP_{p_{i,-}}}{d\varphi du}\right\vert_{u=(p_{i,-}-p_-)/p_-}=\frac{p_{i,-}}{p^2_-} \left.\frac{dP_{p_{i,-}}}{d\varphi du}\right\vert_{u=(p_{i,-}-p_-)/p_-},\\
\frac{dP_{p_-}}{d\varphi dk_-}&=\frac{du}{dk_-}\left.\frac{dP_{p_-}}{d\varphi du}\right\vert_{u=k_-/(p_--k_-)}=\frac{p_-}{(p_--k_-)^2} \left.\frac{dP_{p_-}}{d\varphi du}\right\vert_{u=k_-/(p_--k_-)},\\
\frac{dP_{p_{i,-}}}{d\varphi dk_-}&=\frac{du}{dk_-}\left.\frac{dP_{p_{i,-}}}{d\varphi du}\right\vert_{u=k_-/(p_{i,-}-k_-)}=\frac{p_{i,-}}{(p_{i,-}-k_-)^2} \left.\frac{dP_{p_{i,-}}}{d\varphi du}\right\vert_{u=k_-/(p_{i,-}-k_-)},
\end{align}
can be employed to calculate the phase evolution of the electron distribution $n_{e^-}(\varphi,p_-)$ and of the photon distribution $n_\gamma(\varphi,k_-)$. This method provides a correct treatment of the incoherent multi-photon emission and thus gives the possibility of taking into account RR not only in the classical but also in the quantum regime at moderate values of $\chi^\ast$ (see also the Appendix B and \cite{DiPiazza:2010mv}). Note that the integral of \Eqref{Kinetic1} over all momenta vanishes, which is in agreement with the conservation of the total number of particles. Also, by multiplying the kinetic equation of the electron and the photon distributions by $p_-$ and $k_-$, respectively, and then integrating over all $p_-$ and $k_-$, one obtains that
\begin{equation}
\label{EnMoConserv}
\frac{\partial}{\partial\varphi}\left[\int_0^\infty\, dp_-\,n_{e^-}(\varphi,p_-)p_-+\int_0^\infty\, dk_-\,n_\gamma(\varphi,k_-)k_-\right]=0,
\end{equation}
expressing the conservation of the total energy minus the total longitudinal momentum. 

In the classical limit where quantum recoil effects are negligible, i.e., at $\chi^\ast\ll1$, Eq. (\ref{Kinetic1}) can be expanded in terms of $\chi(\varphi,p_-)$, as shown in \cite{Neitz2013}. Considering only the terms proportional to $\chi^2(\varphi,p_-)$ Eq. (\ref{Kinetic1}), becomes the continuity equation
\begin{equation}
\label{Kinetic_cl}
\frac{\partial n_{e^-}(\varphi,p_-)}{\partial\varphi}=-\frac{\partial}{\partial p_-}\left[n_{e^-}(\varphi,p_-)\frac{d p_-}{d\varphi}\right],
\end{equation}
where
\begin{equation}
\label{pminLL}
\frac{d p_-}{d\varphi}=-\frac{I_{cl}(\varphi)}{\omega_0},
\end{equation}
with
\begin{equation}
I_{cl}(\varphi)=\frac{2}{3}\alpha m^2 \chi^2(\varphi,p_-)
\end{equation}
being the classical intensity of radiation. \Eqref{pminLL} is exactly the classical single-particle equation resulting from the LL equation \cite{DiPiazza:2008}. The analytical solution of \Eqref{pminLL} found in \cite{DiPiazza:2008},
\begin{equation}
\label{LLansol}
p_-(\varphi)=\frac{p_{0,-}}{h(\varphi)},
\end{equation}
where
\begin{equation}
\label{hfunc}
h(\varphi)=1+\frac{2}{3}\alpha\frac{p_{0,-}}{\omega_0}\frac{E_0^2}{E^2_{\text{cr}}}\int_0^\varphi d\phi f^2(\phi),
\end{equation}
shows that the final value $p_-(\infty)$ depends on the plane-wave's parameters only through the total fluence
\begin{equation}
\Phi=\frac{E_0^2}{\omega_0}\int_0^\infty d\phi f^2(\phi).
\end{equation}
Since the analytical solution for the single-particle equation is known, the method of characteristics can be employed to determine the solution of Eq. (\ref{Kinetic_cl}). In fact, if the initial distribution $n_{e^-}(0,p_-)$ is given by the Gaussian distribution
\begin{equation}
\label{indistr}
n_{e^-}(0,p_-)=\frac{N_{e^-}}{\sqrt{\pi/2}\sigma_{p_-}[1+\text{erf}(p_-^\ast/\sqrt{2}\sigma_{p_-})]}\exp\left[-\frac{(p_--p_-^\ast)^2}{2\sigma_{p_-}^2}\right], 
\end{equation}
where $p_-^\ast$ is the average value of $p_-$, $\sigma_{p_-}$ is the standard deviation \cite{Footnote}, $N_{e^-}$ is the total number of electrons and $\text{erf}(x)$ is the error function, then the solution of Eq. (\ref{Kinetic_cl}) reads (see \cite{Neitz2013})
\begin{equation}
\label{ansoldistr}
n_{e^-}(\varphi,p_-)=\frac{N_{e^-}}{\sqrt{\pi/2}\sigma_{p_-}[1+\text{erf}(p_-^\ast/\sqrt{2}\sigma_{p_-})] g^2(\varphi,p_-)}\exp\left\{-\frac{1}{2\sigma_{p_-}^2}\left[\frac{p_-}{g(\varphi,p_-)}-p_-^\ast\right]^2\right\}, 
\end{equation}
with
\begin{equation}
\label{gfunc}
g(\varphi,p_-)=1- \frac{2}{3}\alpha\frac{p_-}{\omega_0}\frac{E^2_0}{E_{\text{cr}}^2}\int_0^\varphi d\phi f^2(\phi).
\end{equation}
Since $p_{0,-}$ in \Eqref{LLansol} is positive for finite values of $p_{0,y}$ and $p_{0,-}\to 0$ only at $p_y\to +\infty$, the function $g(\varphi,p_-)$ must be non-negative for all values of $\varphi$ and the equation $g(\varphi,p_{-,\text{max}})=0$ fixes the maximum value $p_{-,\text{max}}=p_{-,\text{max}}(\varphi)$ allowed for the variable $p_-$ at each $\varphi$. Equation (\ref{ansoldistr}) also indicates that the final electron distribution depends on the plane-wave's pulse shape only via the fluence.

Now, the leading quantum corrections to the classical kinetic equation (\ref{Kinetic_cl}) were shown to change the structure of the latter from a Liouville-like to a Fokker-Planck-like equation \cite{Neitz2013}. This implies that the corresponding single-particle equation becomes a stochastic differential equation. In turn, the full quantum calculations predict a broadening of the electron energy distribution explained by the stochastic nature of photon emission \cite{Neitz2013}, whereas in the classical regime, RR was shown to reduce the energy width of laser-produced electron \cite{Zhidkov_2002} and ion bunches \cite{Naumova_2009,Chen_2010,Tamburini_2010,Tamburini_2011}.

\section{Pulse-shape effects}
\label{Shape}
In this section, we study the influence of the laser pulse form $f(\varphi)$ on the electron and photon distributions, by solving numerically the kinetic equations (\ref{Kinetic1})-(\ref{Kinetic2}). Envisaging an experimental investigation of these effects, we focus on laser pulses at a given pulse fluence that can be modified in pulse shape and pulse duration via the various available pulse shaping techniques (see, e.g., \cite{Winterfeldt:2008} for a discussion of pulse shaping techniques in the context of High Harmonic Generation). We mention that in all our numerical calculations we applied a finite difference method. In the following simulations we will always assume a central angular frequency of the laser field corresponding to the laser photon energy $\omega_0=1.55\,\text{eV}$.

\begin{figure}
\includegraphics[width=\textwidth]{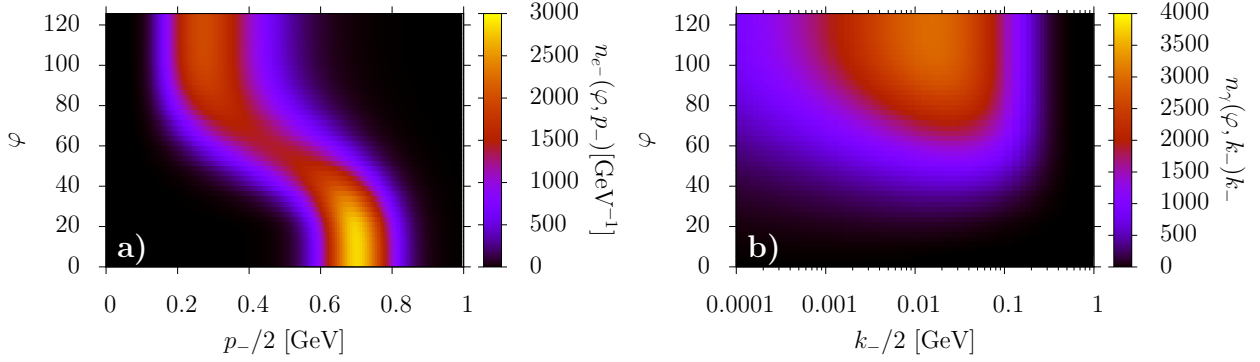} 
\caption{(color online) Phase evolution of the electron distribution (part a)) as a function of $p_-/2\approx\varepsilon$ and the photon spectrum (part b)) as a function of $k_-/2\approx\omega$ for the shape function $f_1(\varphi)$.}
\label{Interv3}
\end{figure}

\begin{figure}
\includegraphics[width=\textwidth]{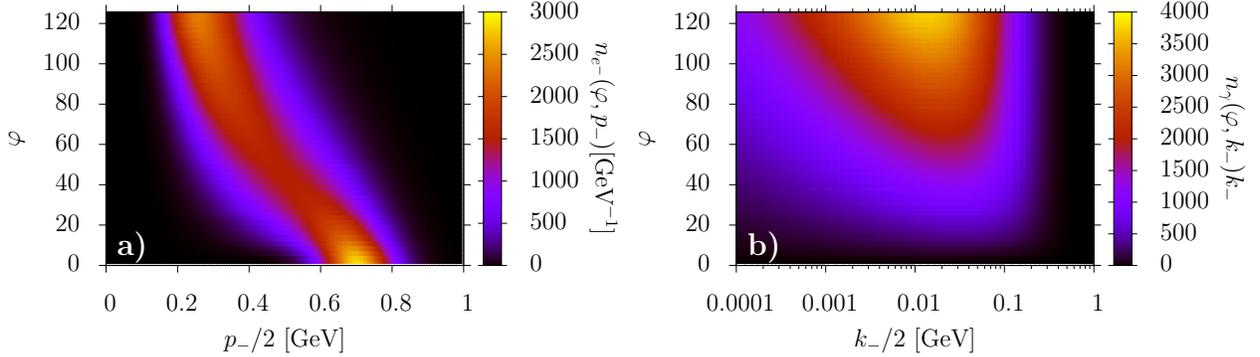}
\caption{(color online) Phase evolution of the electron distribution (part a)) as a function of $p_-/2\approx\varepsilon$ and the photon spectrum (part b)) as a function of $k_-/2\approx\omega$ for the shape function $f_2(\varphi)$.}
\label{Interv4}
\end{figure}

Firstly, we consider two different shapes of the laser pulse at a given pulse fluence and pulse duration. The incoming electron beam is described by the Gaussian beam in Eq. (\ref{indistr}) with $N_{e^-}=1000$, $p_{-}^\ast=1.4\,\text{GeV}$, corresponding to an average energy of $\varepsilon^\ast\approx700\,\text{MeV}$ and with $\sigma_{p_-}=0.14\,\text{GeV}$. We considered two pulses of 20 cycles (final phase $\varphi_f=40\pi$), the first one described by the function $f_1(\varphi)=\sin(\varphi)\sin^2(\varphi/40)$, with a peak intensity of $I_{0,1}=10^{22}\,\text{W/$\text{cm}^2$}$, and the second one by the function 
\begin{equation}
f_2(\varphi)=
\begin{cases}
\sin(\varphi)\sin^2\left(\frac{\varphi}{4}\right)& \text{if } \varphi\in[0,2\pi]\\
\sin(\varphi)& \text{if } \varphi\in[2\pi,38\pi]\\
\sin(\varphi)\sin^2\left(\frac{\varphi-36\pi}{4}\right) & \text{if } \varphi\in[38\pi,40\pi].
\end{cases}
\end{equation}
The peak intensity for the second pulse is $I_{0,2}=4\times10^{21}\,\text{W/$\text{cm}^2$}$, which leads to the same fluence $\Phi=1.3\times 10^9\;\text{J/cm$^2$}$ for both pulses. The difference between the two pulses is that for the second pulse the intensity increases and decreases steeply over just one laser cycle instead of the smooth alteration over the whole pulse length in case of the first pulse. For the above physical scenario we have that the relativistic parameter $\xi$ and the quantum nonlinearity parameter $\chi^\ast$ are $\xi_1=48$ and $\chi_1^\ast=0.40$ for the pulse shape $f_1(\varphi)$, and $\xi_2=31$ and $\chi_2^\ast=0.25$ for the pulse shape $f_2(\varphi)$. Thus, since in both cases it is $\xi\gg 1$, the quasi static approximation can be applied. We note that for the above numerical parameters we are slightly below the quantum radiation dominated regime, which is characterized by the conditions $R_Q=\alpha\xi\sim 1$ and $\chi^\ast\lesssim 1$ \cite{DiPiazza:2010mv}. The evolution of the 
electron distributions $n_{e^-}(\varphi,p_-)$ and the photon spectra $n_\gamma(\varphi,k_-)k_-$ is shown in \Figref{Interv3} for the pulse shape $f_1(\varphi)$ and in \Figref{Interv4} for the pulse shape $f_2(\varphi)$.
In \Figref{Interv3}b) it can be seen, that for the pulse shape $f_1(\varphi)$ the electrons emit less energy at the beginning and more at the peak of the pulse. On the contrary, for the pulse shape $f_2(\varphi)$ (see \Figref{Interv4}b)), the emission of photons starts almost immediately, because the laser profile increases to the maximum value only over one cycle. This effect is also visible in the evolution of the electron distributions in \Figref{Interv3}a) and \Figref{Interv4}a).
\begin{figure}
\includegraphics[width=\textwidth]{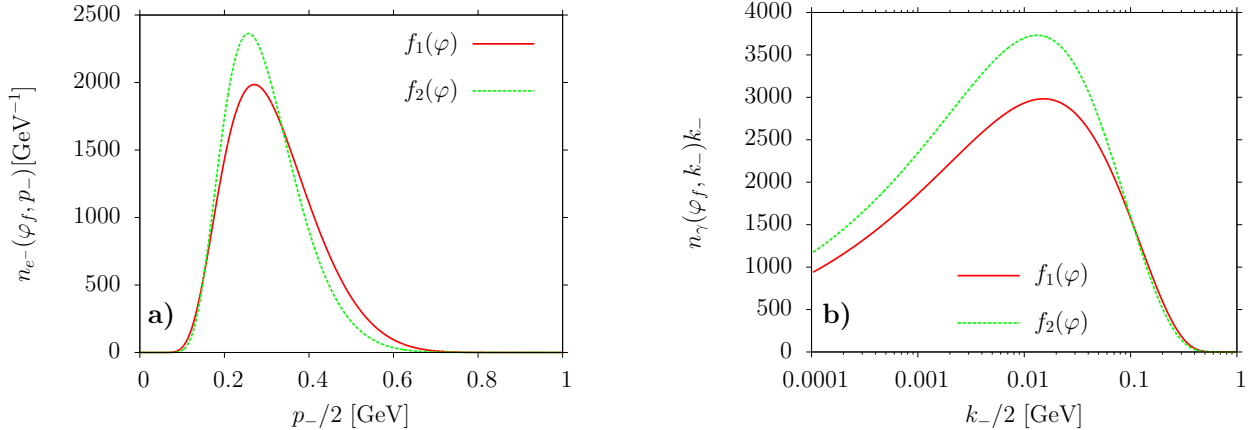}
\caption{(color online) Comparison of the final electron distributions (part a)) as functions of $p_-/2\approx\varepsilon$ and photon spectra (part b)) as functions of $k_-/2\approx\omega$ for the shape functions $f_1(\varphi)$ (solid, red line) and $f_2(\varphi)$ (dashed, green line).}
\label{Interv34}
\end{figure}
Also, since the emission probability increases at higher intensities (see also the final photon spectra in \Figref{Interv34}b)), the photon yield for the pulse shape $f_2(\varphi)$ exceeds the photon yield for the pulse shape $f_1(\varphi)$ due to the longer interaction time at a higher intensity. Thus, the electrons lose more energy in the collision with the laser with the shape function $f_2(\varphi)$. We point out that, since the final electron distribution according to the classical analytical solution in Eq. (\ref{ansoldistr}) depends only on $\Phi$, the differences between the two final electron distributions (see \Figref{Interv34}a)) are due to quantum effects in the interaction. This is in agreement with the values $\chi_1^\ast=0.40$ and $\chi_2^\ast=0.25$ of the typical quantum nonlinearity parameter for the pulse shapes $f_1(\varphi)$ and $f_2(\varphi)$, respectively, which are not significantly smaller than unity. Note that these effects are however smaller than that reported in \cite{Neitz2013}, 
which remains the more prominent signature of quantum RR. We also observe that the average value of the quantity $p_-$ at the end of the interaction of the electron bunch with the laser field ($\sim 500\;\text{MeV}$) is still much larger than the typical transverse momentum $m\xi$ ($\sim 25\;\text{MeV}$ for the pulse shape $f_1(\varphi)$ and $\sim 16\;\text{MeV}$ for the pulse shape $f_2(\varphi)$) (on this respect, see also Eq. (3) in the Supplemental Material of \cite{Neitz2013}). We have ensured that this condition is also fulfilled in the other numerical examples presented below. In addition, even though we have not considered here the dynamics of the beam in the transverse direction, we expect from the analytical solution of the LL equation (see \cite{DiPiazza:2008}) that radiation reaction mainly decreases the transverse momentum of an electron as $p_z\sim m\xi\int_0^\varphi d\phi \,h(\phi)f(\phi)/h(\varphi)\lesssim m\xi$ and $p_x\equiv0$.

\begin{figure}
\includegraphics[width=\textwidth]{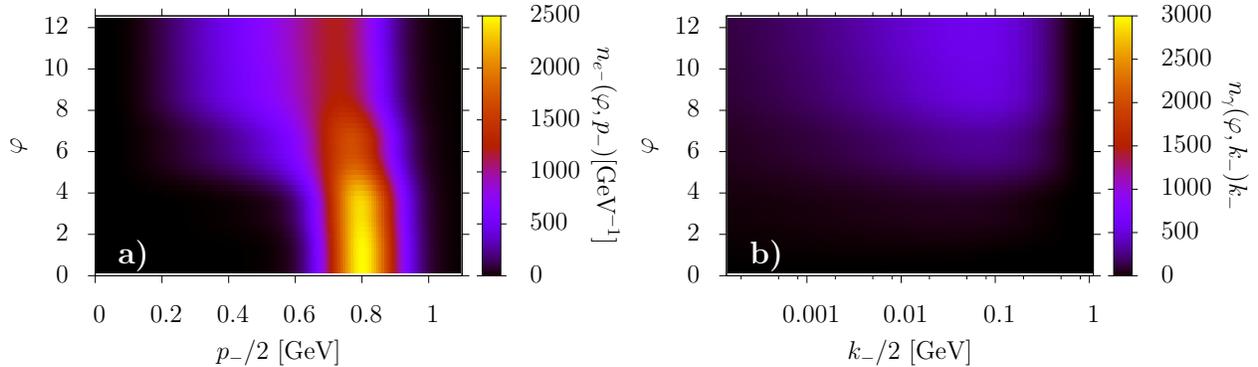}
\caption{(color online) Phase evolution of the electron distribution (part a)) as a function of $p_-/2\approx\varepsilon$ and the photon spectrum (part b)) as a function of $k_-/2\approx\omega$ for the shape function $f_3(\varphi)$.}
\label{Interv5}
\end{figure}

\begin{figure}
\includegraphics[width=\textwidth]{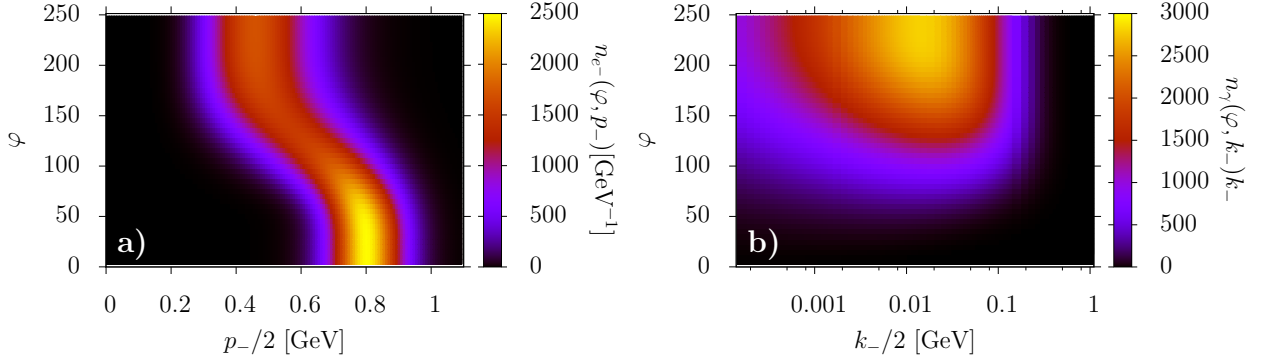}
\caption{(color online) Phase evolution of the electron distribution (part a)) as a function of $p_-/2\approx\varepsilon$ and the photon spectrum (part b)) as a function of $k_-/2\approx\omega$ for the shape function $f_4(\varphi)$.}
\label{Interv7}
\end{figure}

\begin{figure}
\includegraphics[width=\textwidth]{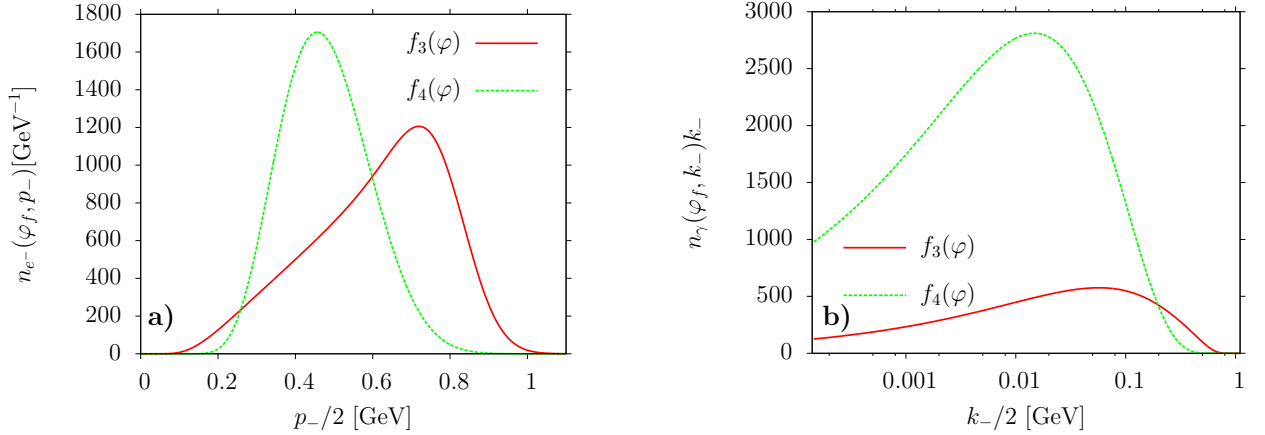}
\caption{(color online) Comparison of the final electron distributions (part a)) as functions of $p_-/2\approx\varepsilon$ and photon spectra (part b)) as functions of $k_-/2\approx\omega$ for the shape functions $f_3(\varphi)$ (solid, red line) and $f_4(\varphi)$ (dashed, green line).} 
\label{Interv57}
\end{figure}

Now, we perform a different comparison of two pulses having the same fluence and the same pulse shape but with different durations and then peak intensities. By keeping the sin$^2$-pulse form, we consider a two-cycle pulse ($\varphi_f=4\pi$), i.e., $f(\varphi)=f_3(\varphi)=\sin(\varphi)\sin^2(\varphi/4)$, with peak intensity $I_{0,3}=4\times10^{22}\,\text{W/$\text{cm}^2$}$ and a 40-cycle pulse ($\varphi_f=80\pi$), i.e., $f(\varphi)=f_4(\varphi)=\sin(\varphi)\sin^2(\varphi/80)$, with peak intensity $I_{0,4}=2\times10^{21}\,\text{W/$\text{cm}^2$}$. For both pulses the fluence $\Phi$ is equal to $5\times 10^8\;\text{J/cm$^2$}$. The initial Gaussian electron beam is centered around $p_{-}^\ast=1.6\,\text{GeV}$, corresponding to an average energy of $\varepsilon^\ast\approx800\,\text{MeV}$, and it has a standard deviation of $\sigma_{p_-}=0.16\,\text{GeV}$. For such an initial electron distribution the quantum nonlinearity parameter $\chi^\ast$ is about unity for the shorter pulse, whereas the relativistic 
parameter is $\xi_3=97$ such that 
$R_Q\approx 0.7$, i.e., the process occurs in the quantum radiation dominated regime. The results of our simulations are shown in \Figref{Interv5} for the pulse shape $f_3(\varphi)$ and in \Figref{Interv7} for the pulse shape $f_4(\varphi)$.
For the two-cycle pulse we observe a completely different phase evolution of the electron distribution than in the previous example. In \Figref{Interv5}a) it is visible that the electron distribution significantly broadens as soon as the laser pulse intensity reaches its maximum and thereby loses its Gaussian shape. Whereas, it can be seen in \Figref{Interv7}a), that the changes in the electron distribution are rather smooth for the longer pulse described by the shape function $f_4(\varphi)$, and the Gaussian shape of the electron distribution is almost conserved. The electron distributions appear to be very sensitive to quantum effects also in the case of the shape function $f_4(\varphi)$, where $\chi_4^\ast\approx 0.2$ ($\xi_4=22$). In fact, the stochasticity of the photon emission cannot be neglected in the quantum regime. In contrast to the classical regime, where the effects of RR are predicted to strongly narrow the electron distributions, the stochastic nature of quantum emission induces a broadening in 
the quantum regime \cite{Neitz2013}. We mention here that we ensured that for values of $\chi^\ast$ smaller than or of the order of $0.01$, the classical and the quantum predictions are found to practically coincide. As in the previous example, we consider the final electron distribution and photon spectrum in more detail, as they can be more relevant from an experimental point of view (see \Figref{Interv57}). 
As we have already pointed out, according to the classical result in \Eqref{ansoldistr} following from the LL equation the final electron distribution does only depend on the fluence of the laser field, which is the same for the two pulse shapes. Thus, the differences between the two electron distributions in \Figref{Interv57}a) arise, as in the previous example, due to quantum effects. We also note that the photon spectrum has its maximum at lower energies for the 40-cycle pulse, and its yield is much higher than for the two-cycle pulse, due to the longer interaction even though at lower laser intensity. 
Finally, we observe that for the two-cycle pulse, where quantum effects are larger, the photon spectrum is peaked at $k_-^\ast\approx 0.2\,\text{GeV}$. Since the probability of pair creation is approximately suppressed by $\eta(k_-^\ast)=\exp(-8/3\varkappa^\ast)$, with $\varkappa^\ast=k_-^\ast\chi^\ast/p_-^\ast$ (see \cite{Ritus:1985}), we conclude that for the peak of the photon spectrum $\eta(k_-^\ast)\sim10^{-12}$ and thus that pair production is negligible, as initially assumed. Further, we ensured that numerical calculations including pair production (see Sec. \ref{PairProSec}) lead to the same results. In addition, assuming a nowadays feasible total number of $N_{e^-}=6\times10^8$ electrons (corresponding to a total charge of $Q=100$ pC) \cite{Wang_2013} we give the estimated number of emitted photons $N_\gamma$ for the numerical simulations above. In the case of the shape functions $f_1(\varphi)$ and $f_2(\varphi)$, we obtain $N_\gamma=9.5\times10^9$ and $N_\gamma=1.1\times10^{10}$, 
respectively, whereas in the case of the shape functions $f_3(\varphi)$ and $f_4(\varphi)$ it results $N_\gamma=1.8\times10^9$ and $N_\gamma=8.5\times10^9$, respectively.

\section{Pair production}
\label{PairProSec}
In this section we include the effect, that photons emitted during the interaction of the electron beam and the laser field may create electron-positron pairs by interacting with the laser field itself (see, e.g., \cite{Sokolov:2010am,Elkina:2010up,Nerush:2010fe} for similar studies). As it was pointed out in \cite{Sokolov:2010am}, however, in the present setup radiation-reaction effects, i.e. the fact that for any elementary process (photon emission and pair photo-production) each final particle has a value of the minus-momentum smaller than the initial particle, makes the generation of a QED cascade impossible. Now, we include terms corresponding to pair production in the kinetic equations (\ref{Kinetic1}) and (\ref{Kinetic2}) to investigate the dynamics of this process together with photon emission. Since an intense laser plane-wave field ($\xi\gg1$) is considered, we are allowed to neglect pair-production processes of higher order, e.g., the direct production of a pair by an electron via the emission of 
a virtual photon \cite{Hu_2010,Ilderton_2011}. Here, again all the probabilities are averaged over the initial photon polarization and summed over the final electron and positron spin. The probability that a photon with momentum $k_-$ produces a pair with particles' momenta $p_-$ and $k_--p_-$ per unit phase $\varphi$ and per unit $p_-$ is given by (see \cite{Ritus:1985})
\begin{equation}
 \label{PairProb}
\frac{dP_{k_-}}{d\varphi dp_-}=\frac{\alpha}{\sqrt{3}\pi}\frac{m^2}{\omega_0k_-^2}\left[\frac{k_-^2}{p_-(k_--p_-)}\,
\text{K}_{\frac{2}{3}}\left( \kappa(\varphi,k_-,p_-)\right)-\int_{\kappa(\varphi,k_-,p_-)}^\infty dx\, \text{K}_{\frac{5}{3}}(x) \right],
\end{equation}
where $\kappa(\varphi,k_-,p_-)=2k_-^2/[3p_-(k_--p_-)\varkappa(\varphi,k_-)]$, with $\varkappa(\varphi,k_-)=(k_-/m)|E(\varphi)|/E_\text{cr}$. By including the process of pair production and by introducing the distribution function $n_{e^+}(\varphi,p_-)$ for the created positrons, we obtain the set of kinetic equations (see Eq. (20.1) in Ref. \cite{Baier:1998vh} and also \cite{Sokolov:2010am})
\begin{align}
\label{KineticPairEl}
\frac{\partial n_{e^-}(\varphi,p_-)}{\partial\varphi}=&\int_{p_-}^\infty dp_{i,-}\,n_{e^-}(\varphi,p_{i,-})\frac{dP_{p_{i,-}}}{d\varphi dp_-}-n_{e^-}(\varphi,p_{-})\int_0^{p_-}dk_-\frac{dP_{p_{-}}}{d\varphi dk_-} \nonumber \\
&+\int_{p_-}^\infty dk_-\,n_\gamma(\varphi,k_-)\frac{dP_{k_-}}{d\varphi dp_-}\\
\label{KineticPairPo}
\frac{\partial n_{e^+}(\varphi,p_-)}{\partial\varphi}=&\int_{p_-}^\infty dp_{i,-}\,n_{e^+}(\varphi,p_{i,-})\frac{dP_{p_{i,-}}}{d\varphi dp_-}-n_{e^+}(\varphi,p_{-}) \int_0^{p_-}dk_-\frac{dP_{p_{-}}}{d\varphi dk_-}\nonumber \\
&+\int_{p_-}^\infty dk_-\,n_\gamma(\varphi,k_-)\frac{dP_{k_-}}{d\varphi dp_-}\\
\label{KineticPairPho}
\frac{\partial n_\gamma(\varphi,k_-)}{\partial\varphi}=&\int_{k_-}^\infty dp_{i,-}\, \left[n_{e^-}(\varphi,p_{i,-})+n_{e^+}(\varphi,p_{i,-})\right]\frac{dP_{p_{i,-}}}{d\varphi dk_-} \nonumber \\
&-n_\gamma(\varphi,k_-)\int_0^{k_-}dp_-\,\frac{dP_{k_-}}{d\varphi dp_-}.
\end{align}

Note that the electron distribution function is no longer decoupled from the photon distribution function and thus the final electron distributions will be affected by the evolution of the photon spectrum. Although the total number of particles is no longer conserved, the integral over all momenta of the 
difference of Eq. (\ref{KineticPairEl}) and Eq. (\ref{KineticPairPo}) vanishes, which implies the conservation of the total charge. Further, the conservation of the total energy minus the total longitudinal momentum is ensured by the analogue of Eq. (\ref{EnMoConserv}) 
\begin{equation}
\label{EnMoConservPair}
\frac{\partial}{\partial\varphi}\left[\int_0^\infty\, dp_-\,n_{e^-}(\varphi,p_-)p_-+\int_0^\infty\, dp_-\,n_{e^+}(\varphi,p_-)p_-+\int_0^\infty\, dk_-\,n_\gamma(\varphi,k_-)k_-\right]=0.
\end{equation}

In order to investigate the dynamics of the created particles and how the pair creation process affects the dynamics of the electrons and photons in the regime $\chi^\ast>1$, we consider different numerical examples below.  Note that the quantum nonlinearity parameter cannot be increased to arbitrary values in our approach, since the energy loss of the electrons due to photon emission would imply a violation of the validity-condition $p_-^\ast\gg m\xi$ of our approach during the laser-particles interaction. However, we ensured that all our approximations are valid throughout the entire numerical simulations. 

Firstly, we consider that a 20-cycle sin$^2$-pulse, i.e., $f(\varphi)=f_5(\varphi)=\sin(\varphi)\sin^2(\varphi/40)$, with laser peak intensity $I_{0,5}=4.2\times10^{21}\,\text{W/$\text{cm}^2$}$ collides with an initially Gaussian electron distribution with $p_-^\ast=100\,\text{GeV}\,(\varepsilon^\ast\approx 50\,\text{GeV})$ and $\sigma_{p_-}=10\,\text{GeV}$ that is normalized to unity. These numerical values correspond to the relativistic parameter $\xi_{5}=31$ and to the 
quantum nonlinearity parameter $\chi_{5}^\ast=19$.
\begin{figure}
\includegraphics[width=\textwidth]{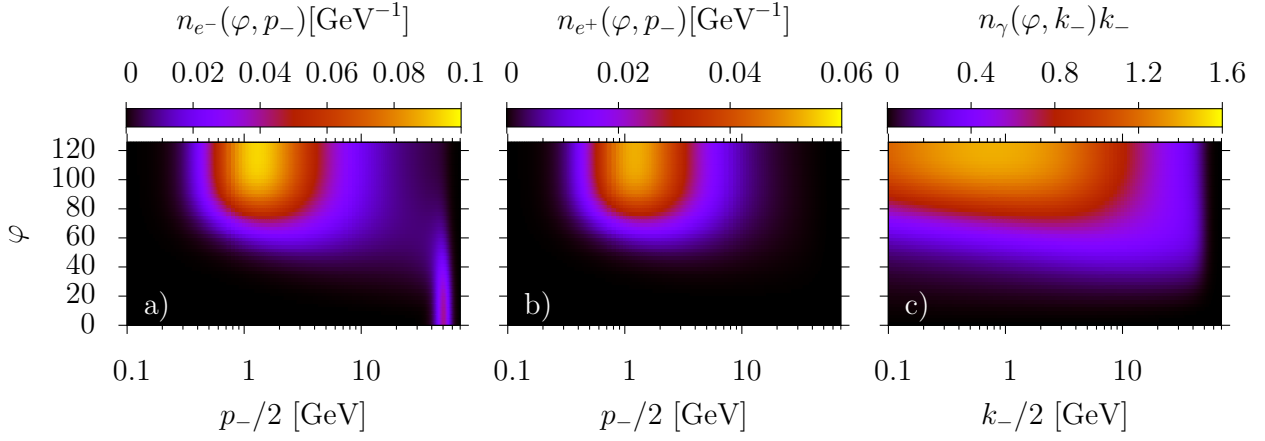}
\caption{(color online) Phase evolution of the electron (part a)) and positron distribution (part b)) as functions of $p_-/2\approx\varepsilon$ and of the photon spectrum (part c)) as a function of $k_-/2\approx\omega$ for the shape function $f_5(\varphi)$.} 
\label{Pair3d}
\end{figure}
The evolutions of the electron distribution, of the positron distribution and of the photon spectrum are shown in \Figref{Pair3d}. In \Figref{Pair3d}a) it can be seen that during the interaction with the laser pulse the electrons lose a large amount of their initial momentum due to the emission of photons. As expected, the so-generated photons have a sufficiently high energy to produce electron-positron pairs. This implies a decrease of the yield of high-energetic photons in the final photon spectrum and, of course, an increases the number of charged particles. In this case, the ratio of the final and the initial number of electrons is approximately $1.56$ corresponding to a growth of the number of electrons by more than 50\%. Mainly the energy of the created particles is much smaller than the initial energy of the electrons (see \Figref{Pair3d}b)), which can be explained by the fact that emitted photons must have a smaller energy than the emitting electrons and that, in addition, the energy of these photons 
is split up into two particles in the pair-production process. Since the evolution of the distribution functions shown in \Figref{Pair3d} includes already the coupled dynamics of electrons, positrons and photons, it is not directly evident how radiation and pair-production processes affect the evolution of each distribution function. In order to gain a deeper understanding of the interplay of the particles, we simulated the same collision process as before but artificially switched off once pair creation and then the radiation of the created positrons. The final distributions of these calculations are shown in \Figref{Pairchar}.
\begin{figure}
\includegraphics[width=\textwidth]{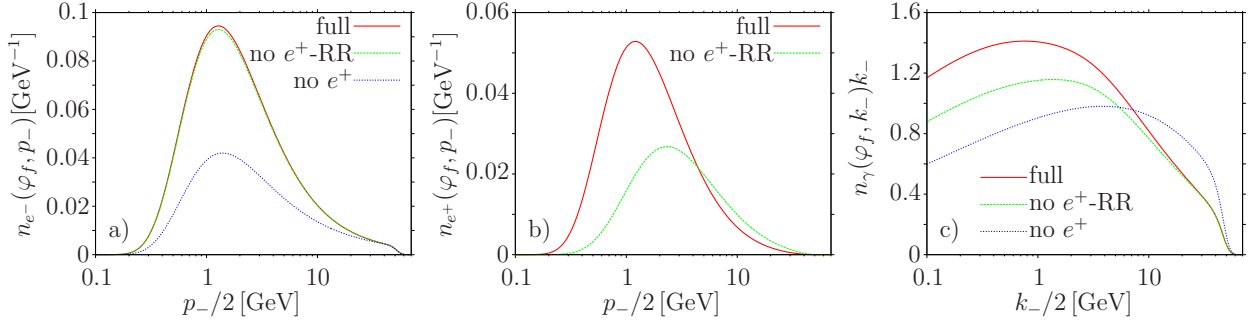}
\caption{(color online) Comparison of the final electron (part a)) and positron distributions (part b)) as functions of $p_-/2\approx\varepsilon$ and photon spectra (part c)) as functions of $k_-/2\approx\omega$ for the full kinetic approach (solid, red line), without the radiation of positrons (dashed, green line) and without pair production (short dashed, blue line) for the initial Gaussian electron distribution with $p_-^\ast=100\,\text{GeV}\,(\varepsilon^\ast\approx 50\,\text{GeV})$ and the shape function $f_5(\varphi)$.} 
\label{Pairchar}
\end{figure}
As, by construction, the positron distribution vanishes identically without the inclusion of pair production,  \Figref{Pairchar}b) only shows the final distribution functions for the full dynamics and the one where radiation by positrons is excluded. Further, it can be seen in \Figref{Pairchar}a) and \Figref{Pairchar}c) that the inclusion of pair creation drastically changes the final electron distribution and photon spectrum. The high-energy part of the photon spectrum is significantly decreased, as in the pair creation process the photon is transformed into a pair of charged particles also corresponding, in turn, to an increase in the low-energy part of the final energy distribution of the electrons. Whereas, the electron distribution at very high energies is not altered, because the produced particles are mostly created at smaller energies. Moreover, the photon gain is enlarged for smaller photon energies, due to the additional emission of the increased number of charged particles. In fact, the inclusion 
of RR for positrons shifts the final energy distribution to lower energies (see \Figref{Pairchar}b)), i.e., the created particles are able to interact with the laser background field after the pair production and in turn will emit photons. In accordance with the argument above, this results in an enlarged photon gain for lower photon energies. Furthermore, the radiation emitted by the positrons barely affects the evolution of the electrons leading to a slightly higher energy spectrum (see \Figref{Pairchar}a)). In this case, the produced positrons had a sufficiently high energy to emit photons that were enabled to again create a small number of pairs during the remaining interaction time with the laser pulse.

Now, we want to examine the influences of the initial energy of the electrons and the laser peak intensities on the final electron and positron distributions and on the photon spectrum. As in Sec. \ref{Shape}, we will again consider laser pulses at a given pulse fluence that can experimentally be modified via pulse shaping. Therefore, we consider two initially Gaussian electron distributions the first as before with $p_-^\ast=100\,\text{GeV}\,(\varepsilon^\ast\approx 50\,\text{GeV})$ and $\sigma_{p_-}=10\,\text{GeV}$ and the second with $p_-^\ast=10\,\text{GeV}\,(\varepsilon^\ast\approx 5\,\text{GeV})$ and $\sigma_{p_-}=1\,\text{GeV}$, which are both normalized to unity. We consider these two electron distributions to collide with three different laser pulses that have the same fluence $\Phi=5.2\times 10^8\;\text{J/cm$^2$}$ and a sin$^2$-pulse form, i.e., $f(\varphi)=\sin(\varphi)\sin^2(\varphi/2N_L)$, but differ in the number of laser cycles $N_L$ and the laser peak intensities. Thus, we have chosen laser 
peak intensities of $I_{0,5}=4.2\times10^{21}\,\text{W/$\text{cm}^2$}$ for $f_{5}(\varphi)=\sin(\varphi)\sin^2(\varphi/40)$ ($\varphi_f=40\pi$, $\xi_{5}=31$ and 
$\chi_{5}^\ast=19$), $I_{0,6}=1.7\times10^{22}\,\text{W/$\text{cm}^2$}$ for $f_6(\varphi)=\sin(\varphi)\sin^2(\varphi/10)$ ($\varphi_f=10\pi$, $\xi_6=63$ and $\chi_6^\ast=37$) and $I_{0,7}=10^{23}\,\text{W/$\text{cm}^2$}$ for $f_{7}(\varphi)=\sin(\varphi)\sin^2(\varphi/2)$ ($\varphi_f=2\pi$, $\xi_{7}=153$ and $\chi_{7}^\ast=91$) corresponding to pulse durations of $60\;\text{fs}$, $15\;\text{fs}$ and $3\;\text{fs}$, respectively.
\begin{figure}
\includegraphics[width=\textwidth]{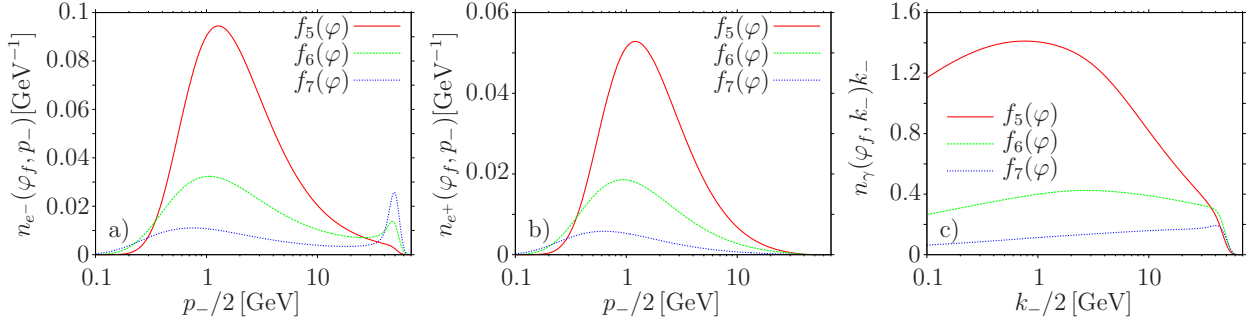}
\caption{(color online) Comparison of the final electron (part a)) and positron distributions (part b)) as functions of $p_-/2\approx\varepsilon$ and photon spectra (part c)) as functions of $k_-/2\approx\omega$ for the shape functions $f_5(\varphi)$ (solid, red line), $f_6(\varphi)$ (dashed, green line) and $f_{7}(\varphi)$ (short dashed, blue line) for the initial Gaussian electron distribution with $p_-^\ast=100\,\text{GeV}\,(\varepsilon^\ast\approx 50\,\text{GeV})$.} 
\label{Pairnbfin}
\end{figure}
In \Figref{Pairnbfin} the final electron and positron distributions as well as the photon spectra are shown for the three collisions with the first electron distribution. In agreement with our previous results, the photon yield is higher for the 20-cycle pulse than for both the five-cycle pulse and the one-cycle pulse. The reason again is the longer interaction time of the electrons with the laser field, although the laser peak intensity is smaller. Since the initial energy of the electrons and the laser intensities are sufficiently high to produce many high-energetic photons during the interaction, also the number of pairs produced by the longer pulses exceed the number produced by the one-cycle pulse (see \Figref{Pairnbfin}). In fact, the ratio of the final and the initial electron number is reduced to $1.06$ for the one-cycle pulse, even though in comparison with the 20-cycle pulse the laser peak intensity is larger by a factor of $24$. In prospect of an experimental investigation, an electron 
bunch with typical charge $Q\approx100$ pC corresponding to a total initial number of $N_{e^-}=6\times 10^8$ electrons is considered in order to achieve estimates for the final number of produced positrons $N_{e^+}$ and photons $N_\gamma$. From our numerical simulations, we conclude that for the shape function $f_5(\varphi)$ it is $N_{e^+}=3.3\times10^8$ and $N_\gamma=5.5\times10^9$, whereas for $f_6(\varphi)$ it results $N_{e^+}=1.3\times10^8$ and $N_\gamma=1.7\times10^9$ and for $f_6(\varphi)$ we obtain $N_{e^+}=3.4\times10^7$ and $N_\gamma=5.5\times10^8$.

\begin{figure}
\includegraphics[width=\textwidth]{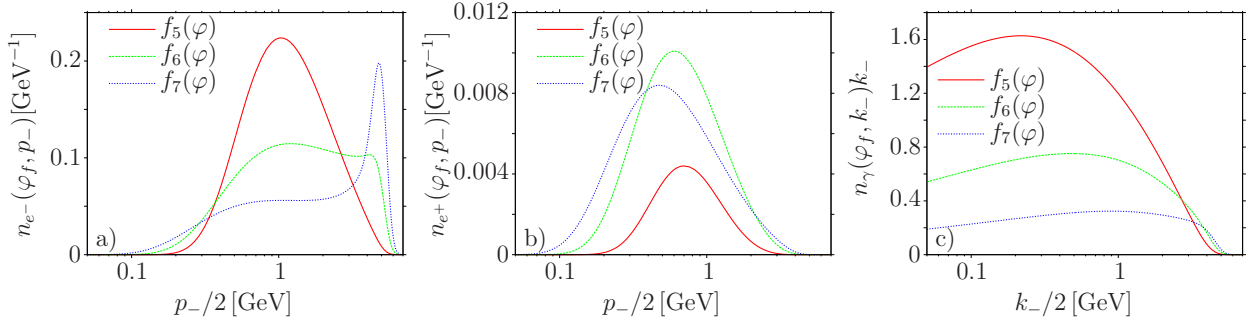}
\caption{(color online) Comparison of the final electron (part a)) and positron distributions (part b)) as functions of $p_-/2\approx\varepsilon$ and photon spectra (part c)) as functions of $k_-/2\approx\omega$ for the shape functions $f_5(\varphi)$ (solid, red line), $f_6(\varphi)$ (dashed, green line) and $f_{7}(\varphi)$ (short dashed, blue line) for the initial Gaussian electron distribution with $p_-^\ast=10\,\text{GeV}\,(\varepsilon^\ast\approx 5\,\text{GeV})$.} 
\label{Pairnfin}
\end{figure}
Since for the second electron distribution centered at $p_-^\ast=10\,\text{GeV}$ the value $\chi^\ast$ is reduced by a factor of $10$, also the value $\varkappa^\ast$ will be decreased and quantum effects and pair production are expected to be less prominent. The final distribution functions for the collision of this electron distribution and the aforementioned three laser pulses are shown in \Figref{Pairnfin}. As in the simulations above the electrons lose most of their energy in the collision with the 20-cycle pulse. On the other hand, the shorter pulses lead to much broader final electron distributions and the initial peak around $p_-^\ast=10\,\text{GeV}\,(\varepsilon^\ast\approx 5\,\text{GeV})$ is still pronounced for the one-cycle pulse. Although this results once more in a much larger photon yield for the long pulse, the spectrum has it maximum at smaller values of $k_-$ than in the previous example (see \Figref{Pairnfin}c)). Thus, the typical value $\varkappa^\ast$ is decreased significantly leading 
to a lower pair-production probability. In fact, the number of produced pairs by the shorter pulses now exceeds the one produced by the 20-cycle pulse. Even though the total gain of photons is larger in the case of the long pulse, the energy of the emitted photons and the laser peak intensity are not sufficiently high to create a large number of pairs. However, for the five-cycle pulse the number of produced particles is slightly higher than for the shortest pulse, i.e., the laser peak intensity is high enough in this case to obtain the beneficial effects of the longer interaction time also in the positron yield (see \Figref{Pairnfin}b)). As in the previous examples, estimations for the final $N_{e^+}$ and $N_\gamma$ are given for an electron bunch with total initial number of $N_{e^-}=6\times 10^8$ electrons. Hence, in the case of $f_5(\varphi)$ one obtains $N_{e^+}=5.8\times10^6$ and $N_\gamma=5.6\times10^9$, whereas in the case of $f_6(\varphi)$ ($f_7(\varphi)$) it results $N_{e^+}=1.6\times10^7$ 
and $N_\gamma=2.5\times10^9$ ($N_{e^+}=1.4\times10^7$ and $N_\gamma=1.0\times10^9$).

Finally, we give an estimation for the laser peak intensities at which electron-positron pairs will be detectable for nowadays available laser accelerated electron beams \cite{Wang_2013}. Therefore, a Gaussian electron beam centered at $p_-^\ast=4\,\text{GeV}\,(\varepsilon^\ast\approx 2\,\text{GeV})$ and $\sigma_{p_-}=0.2\,\text{GeV}$ with again a total number of $N_{e^-}=6\times 10^8$ electrons is considered to collide with 10-cycle sin$^2$-shaped laser pulses with different peak intensities. Considering that the detection of a few tens of positrons is feasible, pair production should be measurable at an intensity of $I_0=1.5\times 10^{21}\,\text{W/$\text{cm}^2$}$, which would lead to a total number of $26$ created positrons. For a slightly lower electron energy $p_-^\ast=2\,\text{GeV}\,(\varepsilon^\ast\approx 1\,\text{GeV})$ the production of $19$ pairs already requires an increased intensity of $I_0=5\times 10^{21}\,\text{W/$\text{cm}^2$}$, indicating that an increase 
in the 
beam energy is more favorable in the studies of pair production.

\section{Conclusion}
In the present paper we have investigated how the pulse shape of the laser and its duration influence the evolution of an electron beam which collides head-on with the laser pulse. By employing a kinetic approach, we have included quantum radiation-reaction effects, in a regime where they mainly stem from multiple incoherent photon emission. In particular, we have investigated the dependence of the final electron distribution and of the final photon spectrum on the pulse shape and on the duration of pulses with the same fluence. By keeping the laser fluence fixed, we ensure that the classical theory of radiation reaction based on the Landau-Lifshitz equation predicts that the final electron spectrum and the total electromagnetic energy emitted by the electrons are the same for different pulses. Thus, possible differences in such observables indicate an interplay of quantum and radiation-reaction effects. Our numerical simulations show that already for $\chi^\ast=0.2\text{-}0.3$ the final electron 
distribution and final photon spectrum are altered by quantum radiation reaction effects. Our results also indicate that these quantum radiation-reaction effects are measurable in principle already at available laser intensities of the order of $10^{22}\;\text{W/$\text{cm}^2$}$ \cite{Yanovsky:2008} and at electron bunch 
energies of the order of $1\;\text{GeV}$ \cite{Wang_2013}.

Furthermore, we have studied how electron-positron pair creation affects the dynamics of the charged particles and photons. The inclusion of electron-positron pair production was shown to significantly decrease the high-energy part of the photon spectra and to increase the low-energy part of the electron and positron distribution, due to the enhanced number of charged particles. Moreover, a weak nonlinear coupling of all three distribution functions became apparent in the fact that the radiation of positrons can (slightly) alter the final electron distribution. Since at a fixed laser fluence the photon gain was found to be higher for longer pulses (though with smaller peak intensity), the creation process is amplified if the initial energy of the electrons is sufficiently high. Finally, we estimated that nowadays high intensity laser facilities \cite{Yanovsky:2008}, as well as high energy electron beams \cite{Wang_2013}, allow in principle for an experimental detection of pair creation in an all-optical 
setup. 

\appendix
\section{On the derivation of the kinetic equation (\ref{Kinetic1})}
In this appendix we revise a step in the derivation of the kinetic equation (\ref{Kinetic1}) of the main text given in the Supplemental Material of \cite{Neitz2013}. We point out, however, that the final form of this equation is unchanged.

In the Supplemental Material of \cite{Neitz2013}, the electron distribution function $f_{e^-}(\phi,T,\bm{r}_\perp,p_-,\bm{p}_\perp)$ has been assumed to have the structure $f_{e^-}(\phi,T,\bm{r}_\perp,p_-,\bm{p}_\perp)=\delta(\bm{p}_\perp)\rho_{e^-}(\phi,T,\bm{r}_\perp,p_-)$ corresponding to a distribution that is ``infinitely'' peaked at $\bm{p}_\perp=\bm{0}$ throughout the whole interaction of the electron beam with the laser plane-wave field. Here, we used the light-cone coordinates $\phi=t-y$, $T=(t+y)/2$ and $\bm{r}_{\perp}=(x,z)$, and the corresponding quantities $p_-=\varepsilon-p_y$, $p_+=(\varepsilon+p_y)/2$ and $\bm{p}_{\perp}=(p_x,p_z)$. However, the specific choice of the delta function is not essential for the derivation of the kinetic equation. We remind that a basic assumption in our approach is that $m\xi\ll \varepsilon$, which allows us already to neglect the terms proportional to the transverse momenta and to $p_+\sim (m^2+\bm{p}^2_{\perp})/\varepsilon^2$ in Eq. (2) in the Supplemental 
Material of \cite{Neitz2013}. Thus, by considering still a well peaked function, e.g., a Gaussian, but also allowing for a finite transverse momentum evolving in phase, we can seek for a solution of the form $f_{e^-}(\phi,T,\bm{r}_\perp,p_-,\bm{p}_\perp)=g_{e^-}(\phi,\bm{p}_\perp)\rho_{e^-}(\phi,T,\bm{r}_\perp,p_-)$. Since $\rho_{e^-}(\phi,T,\bm{r}_{\perp},p_-)$ can be factorized by $\rho_{e^-}(\phi,T,\bm{r}_{\perp},p_-)=f_T(T)f_{\perp}(\bm{r}_{\perp})n_{e^-}(\phi,p_-)$, we employ the modified ansatz in the approximated kinetic equation mentioned in the Supplemental Material of \cite{Neitz2013} and obtain the following equation (which replaces Eq. (6) in the Supplemental Material of \cite{Neitz2013})
\begin{align}
\left[\frac{\partial}{\partial \phi}+eE(\phi)\frac{\partial}{\partial p_z}\right]&g_{e^-}(\phi,\bm{p}_\perp)n_{e^-}(\phi,p_-)=\nonumber\\
&\quad g_{e^-}(\phi,\bm{p}_\perp)\left[\int_0^{\infty} d k_-\,\frac{d P(\phi;p_-+k_-\to p_-)}{d\phi d k_-}n_{e^-}(\phi,p_-+k_-)\right.\\
&\quad\left.-n_{e^-}(\phi,p_-)\int_0^{\infty} d k_-\,\frac{d P(\phi;p_-\to p_--k_-)}{d\phi d k_-}\right]\nonumber.
\end{align}
We recall that in deriving this equation, we have also assumed that the emission of photons does not significantly broaden the electron distribution in the transverse momenta with respect to the broadening already induced by the laser field (i.e. by the Lorentz force). Within our model, this assumption is justified as in the ultra-relativistic regime the electrons mainly emit along the instantaneous propagation direction \cite{Baier:1998vh}, i.e. almost along the negative $y$ direction in the present case (see also the Supplemental Material of \cite{Neitz2013}). For the sake of consistency, in our simplified model, we have neglected the small angular spreading (of the order of $m\xi/\varepsilon\ll 1$) brought about by the Lorentz force as well as by the emission of photons off-axis. Now, the function $g_{e^-}(\phi,\bm{p}_\perp)$ can be chosen to satisfy the Liouville-like equation
\begin{equation}
\label{P_perp}
\left[\frac{\partial}{\partial \phi}+eE(\phi)\frac{\partial}{\partial p_z}\right]g_{e^-}(\phi,\bm{p}_\perp)=0.
\end{equation}
If $g_{e^-}(0,\bm{p}_\perp)$ is a given function $\tilde{g}(\bm{p}_\perp)$ well peaked around $\bm{p}_\perp= \bm{0}$ (in the limiting case, it could also be $\tilde{g}(\bm{p}_\perp)=\delta(\bm{p}_\perp$)), then the solution of Eq. (\ref{P_perp}) is $g_{e^-}(\phi,\bm{p}_\perp)=\tilde{g}(\bm{p}_\perp+e\bm{A}(\phi))$, where $\bm{A}(\phi)=(0,0,-\int_0^\phi d\phi' E(\phi'))$ (see also \Eqref{pPerpEvolv} in the main text). Thus, the evolution of the transverse electron momentum driven by the external plane-wave field decouples from the longitudinal one and we again obtain that the reduced distribution function $n_{e^-}(\phi,p_-)$ satisfies Eq. (7) in the Supplemental Material of \cite{Neitz2013}.

\section{The microscopic approach}
\begin{figure}
\centering
\includegraphics[width=0.7\textwidth]{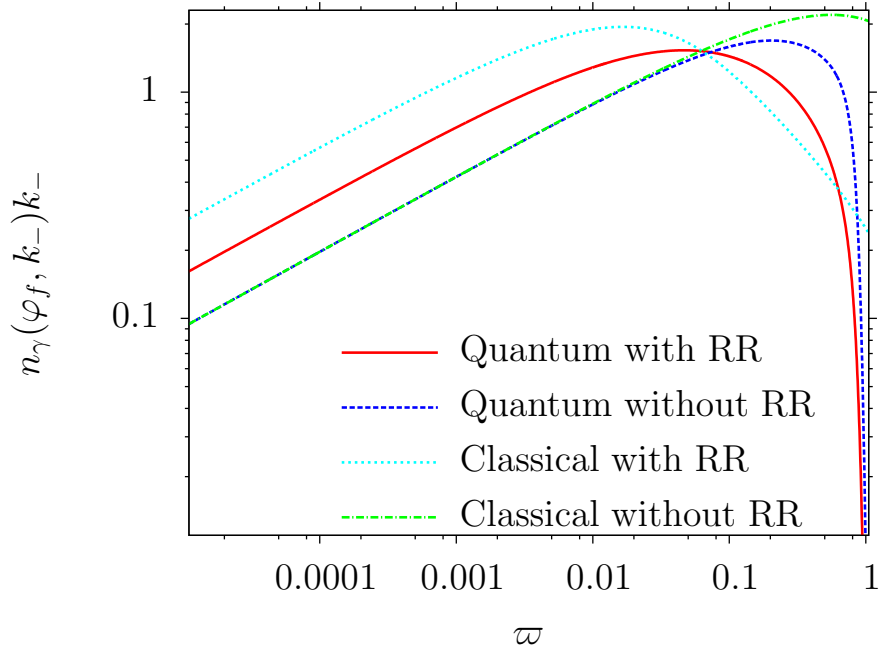}
\caption{(color online) Photon emission spectra: solid, red line (quantum spectrum with RR effects), long dashed, dark blue line (quantum spectrum without RR effects), short dashed, light-blue line (classical spectrum with RR effects) and dot-dashed, green line (classical spectrum without RR effects).}
\label{QClRR}
\end{figure}

In this appendix the results of the so-called microscopic approach developed in \cite{DiPiazza:2010mv} are compared with the ``macroscopic'' kinetic approach employed here. In the microscopic approach the starting point to calculate the photon spectrum is the probability that a single electron emits more than one photon incoherently. The single-photon spectrum is then obtained by essentially integrating over the degrees of freedom of all emitted photons except one. In the microscopic approach one avoids the solution of the kinetic integro-differential equations (\ref{Kinetic1})-(\ref{Kinetic2}). However, the calculation of multi-dimensional integrals is required, with dimensionality increasing with the number of photons emitted. In the numerical example considered in \cite{DiPiazza:2010mv} the spectrum was found to converge after the inclusion of $16$ photons and the corresponding multidimensional integrals were evaluated via the Monte Carlo method. For the sake of comparison with that numerical example, we 
consider a two-cycle sinusoidal pulse, i.e., $f(\varphi)=\sin(\varphi)$, with a peak intensity of $I_0=10^{23}\,\text{W/$\text{cm}^2$}$ and again with $\omega_0=1.55\,\text{eV}$ (note that in \cite{DiPiazza:2010mv} the given value of the intensity was the average one). We center the initial Gaussian electron distribution around $\varepsilon^\ast=1\,\text{GeV}$ (corresponding to the value $p^\ast_-\approx 2\;\text{GeV}$ employed in \cite{DiPiazza:2010mv}), and consider a standard deviation of $\sigma_{p_-}=0.1\,\text{GeV}$. In order to compare with the single-particle approach in \cite{DiPiazza:2010mv} and with the numerical results presented in Fig. 2 there, we fix $N_{e^-}=1$ and we show the photon spectra with respect to the normalized quantity $\varpi=k_-/p^\ast_-$. The resulting final photon spectrum is the quantum photon spectrum including RR effects, i.e., multiple incoherent photon emissions, shown in \Figref{QClRR} as a solid, red line.
The quantum spectrum without RR is obtained by averaging the single-photon emission spectrum (see Eq. (\ref{ToniProb})) with respect to the initial electron distribution. On the other hand, the classical spectrum without RR effects can be obtained by multiplying the single-photon emission probability in \Eqref{ToniProb} with $1+u\approx 1$ and $u\approx k_-/p_-$ by $k_-$, and the classical RR effects can be included by employing the analytical expression of $p_-(\varphi)$ in \Eqref{LLansol} according to the LL equation. A comparison of the quantum spectrum including RR effects in \Figref{QClRR} (solid, red line), with the solid black line in Fig. 2 in \cite{DiPiazza:2010mv}, shows the excellent agreement of the results in the two different approaches.


\begin{thebibliography}{99}
\bibitem{Footnote} This is strictly true if the variable $p_-$ spans from $-\infty$ to $+\infty$. However, the momentum $p_-$ allows only for positive values, and in the case where $p_-^\ast\gg \sigma_{p_-}$ (which will be fulfilled in all our numerical examples) $p_-^\ast$ and $\sigma_{p_-}$ are a good approximation of the average value and of the standard deviation of the distribution, respectively.
\end{thebibliography}
\end{document}